# Planets: Power Laws and Classification

Hector Javier Durand-Manterola

Departamento de Estudios Espaciales, Instituto de Geofísica,
Universidad Nacional Autónoma de México (UNAM)
hdurand_manterola@yahoo.com

Abstract
The study of the interior of the planets requires the knowledge of how certain parameters, as radius and mean density, vary according to the planet mass. The aim of this work is to use known data of the Solar System Planets and Transiting Exoplanets (specifically the radius and mass) to create empirical laws for the planetary radius, mean density, and surface gravity as a function of mass. The method used is to calculate with the available data, the mean density and surface gravity for the planets and adjusts, using the least squares method, a function with respect to the radius-mass, density-mass and surface gravity-mass relations. In the mass interval from $10^{19}$ to $10^{29}$ kg, the planets separate in a natural way into three groups or classes which I called class A, class B and class C. In all these classes and with all the functions (radius, median density and surface gravity) those best fits are power laws.

Introduction
When studying the general laws governing the structure of bodies such as the exoplanets, Jupiter, The Earth, The Moon and Titan, Enceladus and Pluto it is not relevant whether they are a moon, a planet or a dwarf planet. Together could be called planetary bodies, but to save words I will call at all, only planets. So in this work is taken as a planet any body that conforms to the following definition:

"A celestial body, no matter how it has formed, no matter which site or neighborhood is located, which has a mass that is below the threshold mass for thermonuclear fusion of deuterium, and a mass large enough so that its shape (spherical) is determined by a viscous relaxation induced by the rotation and gravity".

From this point of view, for the purpose of this work, will be planets, not just the eight planets accepted by the IAU, but also The Moon, the four Galilean Satellites, Titan; Enceladus or any other spherical satellite, as well as the dwarf planets Pluto, Eris, Ceres or any other bodies within the Kuiper Belt which conforms with the previous description. The exoplanets will also be considered in this class. Although there are no images of them their great mass leads us to think that they should have spheroid shapes determined by gravity and rotation. The above definition coincides essentially with the definition of "planemo" given by Basri (2003), but in this work, I will not use this term because I believe that the word "planet" sooner or later will be used in this sense. For example in Planetary Sciences it is included the study of the moon and other bodies not considered planets by the IAU definition (Melosh, 2011).

One of the main problems, when studying a planet, is finding out how the mass is distributed within the interior (Valencia et al., 2007; Swift et al., 2011). The first step to understanding the distribution within the interior of the planets is to see how the radius, mean density and surface gravity behave as the planet's mass increases. The mass M and radius R of the planets in our Solar System as well as the mass and radius of the transiting extrasolar planets are known, so those parameters can be calculated. Using the data of 26 planets of Solar System (Tholen et al., 2000) and 92 transiting exoplanets (Schneider, 2010), an empirical power law R(M) was obtained for all these bodies. Planets with a mass > $3 \times 10^{25}$kg do not adjust too well to this general radius-mass law and they seem to form two independent groups. For this reason the planets was separated in three classes A, B, and C. Class A is for the planets with a mass < $3 \times 10^{25}$kg, Class B has the ones in the range of $3 \times 10^{25} < M < 1 \times 10^{27}$ and Class C are the planets with a mass > $10^{27}$kg.

Taking separately each class, each one satisfies a different radius-mass power law. The slope in the power law of class A is less pronounced than the one of the general power law. The power law of class B has a greater slope and class C has no slope, showing that in this class the radius remains constant when the mass increases. I also got power laws for the mean density of the three classes of planets studied. Here also we see, more clearly, the separation of the planets into three classes. In classes A and C the density increases when mass increases, but with class B the density decreases when mass increases. Power laws for surface gravity were also obtained for these three groups of planets and here again clearly shows the separation into three classes.

2 Power Laws and Classes

2.1 *General radius-mass power law*

To understand the behavior of a planet there is one question that must be answered, what is the relationship between the planetary radius R and the planetary mass M? For the 118 planets considered, the R (M) equation obtained has the following power law:

$$R = 0.0342 M^{0.3428} \tag{1}$$

To obtain function (1) the following data was used: the radius of 26 Solar System planets (Tholen et al., 2000), the mass of the same 26 Solar System planets (Tholen et al., 2000), the radius of 92 transiting exoplanets (Schneider, 2010), and the mass of these same 92 transiting exoplanets (Schneider, 2010). Schneider's data is presented as Jovian mass and Jovian radius, so it must be converted into kilograms and meters. With the least squares method a power curve was adjusted to the data, this is equation (1) (figure 1).

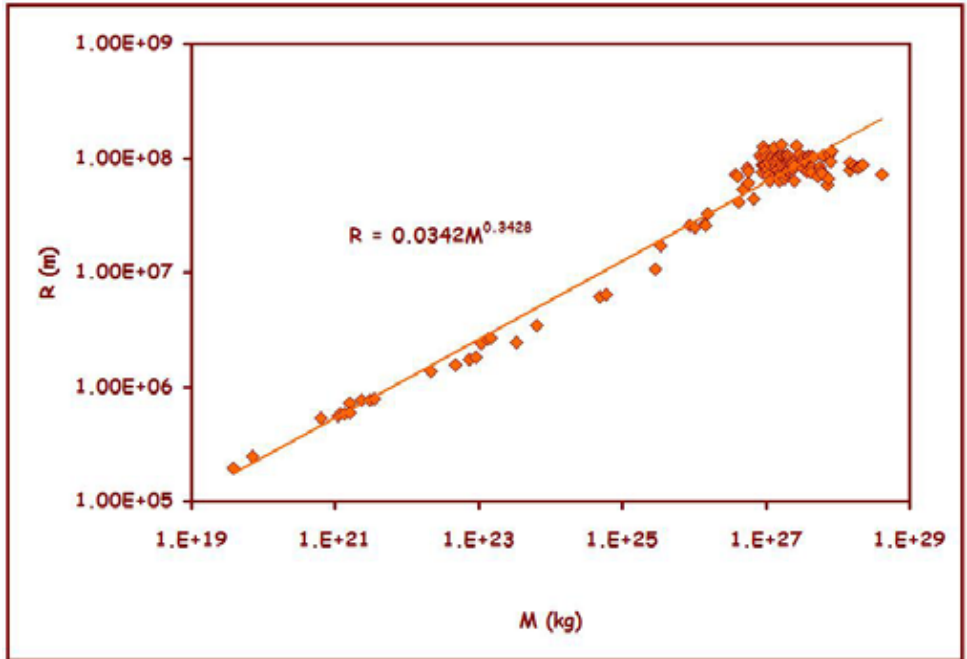

Figure 1 Log-log graph of the planetary radius vs. the planetary mass. The orange line is the adjusted power function (equation 1).

2.2 Separation into Classes

In figure 1 we can see that the planets with a mass smaller than $3\times10^{25}$kg seem to follow a power law much softer than equation (1). The ones in the range from $3\times10^{25}$kg to $10^{27}$kg seem to have a harder power law, and the planets with a mass greater than $10^{27}$kg have a slope near zero. Considering this we can separate the planets into three classes, A, B and C.

a) The power law which fits class A ($M < 5\times10^{25}$kg) is:

$$R = 0.5501 M^{0.2858} \qquad (2a)$$

b) The power law which fits class B ($5 \times 10^{25}$ kg $< M < 10^{27}$ kg) is:

$$R = 7 \times 10^{-8} M^{0.5609} \tag{2b}$$

c) And the power law which fits class C ($M > 10^{27}$ kg) is:

$$R = 4 \times 10^{8} M^{-0.0241} \tag{2c}$$

To obtain equations (2a), (2b) and (2c), it was made a least squares fitting using the same data as in figure 1, but separated in the proposed intervals. The data and the fitted power laws appear in figure 2.

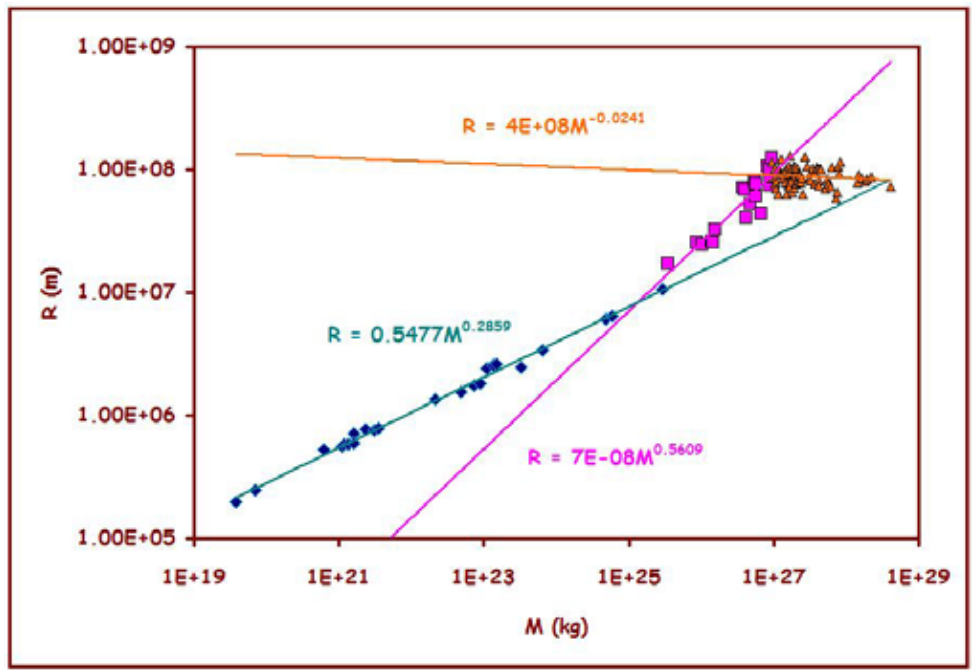

Figure 2 Log-log graph of the planetary radius vs. the planetary mass and fitted power functions for the three mass intervals $M < 5 \times 10^{25}$ kg (blue diamonds), $5 \times 10^{25}$ kg $< M < 10^{27}$ kg (pink squares), and $M > 10^{27}$ kg (orange triangles).

2.3 Mean density in the three classes of planets

When the planetary mean density is graphed against their mass, we see more clearly the separation of the planets into three classes

(figure 3). The density for the three classes of planets also follows a power law as is the case of the radius.

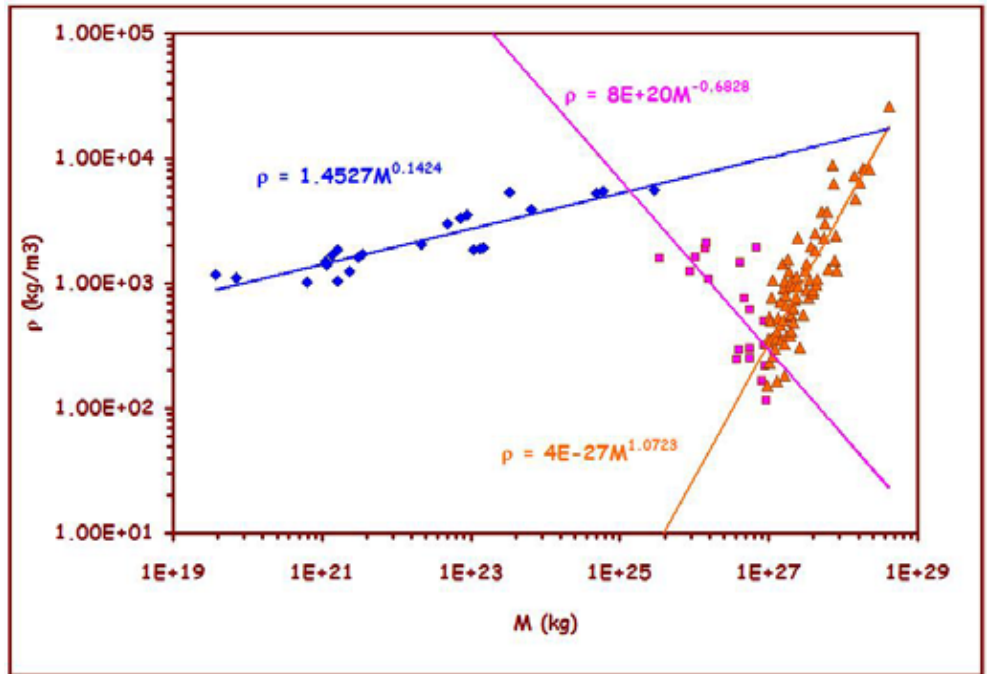

Figure 3 Log-log graph of the mean planetary density vs. the planetary mass and fitted power functions for the three mass intervals M<5x10$^{25}$kg (blue diamonds), 5x10$^{25}$kg<M<10$^{27}$kg (pink squares), and M>10$^{27}$kg (orange trangles).

In class A the density fits to the curve

$$\rho_A = 1.4527 M^{0.1424} \tag{3a}$$

In class B the density fits to the curve

$$\rho_B = 8 \times 10^{20} M^{-0.6828} \tag{3b}$$

And in class C the density fits to the curve

$$\rho_C = 4 \times 10^{-27} M^{1.0723} \tag{3c}$$

arXiv: November 16, 2011

Equations (3a), (3b) and (3c) were obtained as follows: Assuming all the planets spherical, the mean density was calculated and these data were fitted power laws by the least squares method using the same interval of mass as in paragraph 2.2.

2.4 Surface gravity
Planetary surface gravity also follows three different power laws for the three classes of planets (figure 4).
In class A the gravity g(M) is given by

$$g(M) = 2 \times 10^{-10} M^{0.4282} \tag{4a}$$

In class B the gravity g(M) is given by

$$g(M) = 14937 M^{-0.1219} \tag{4b}$$

And in class C the gravity g(M) is given by

$$g(M) = 4 \times 10^{-28} M^{1.0482} \tag{4c}$$

Equations (4a), (4b) and (4c) were obtained as follows: it was calculated the surface gravity for each planet using the gravity law and dividing by the mass of a body placed on the surface of the planet with mass $M_p$ and radius R, then we obtain

$$g = G \frac{M_p}{R^2} \tag{5}$$

From equation (5) and using the radius and the mass of every planet, the surface gravity g is obtained. By the method of least squares were fitted potential curves for the same mass intervals of paragraph 2.2.

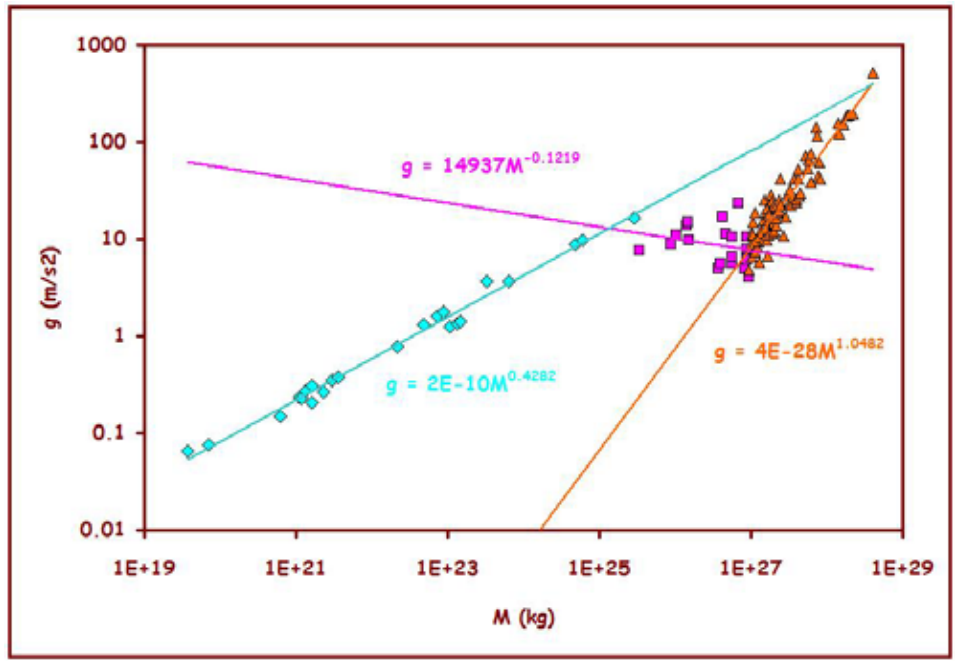

Figure 4 Surface gravity acceleration for planets separated in the three classes and their fitted curves.

3 Discussion

The critical correlation coefficient ($R_C$) is the value that must exceed a coefficient of correlation R in a sample of size N to be statistically significant at alpha level considered.

Correlation coefficients for the 9 equations can be found in table 1. For every class, at least two correlation coefficients are greater than the critical coefficient and this allows us to say that there is a correlation between the data and thus classify the planets in three classes A, B and C.

Table 1

Correlation coefficients for all the power laws obtained and the critical correlation coefficient $R_C(N)$ at 0.005 alpha level for N planets.

|   | N  | $R_C$ [1] | R(M)  | ρ(M)  | g(M)  |
|---|----|-----------|-------|-------|-------|
| A | 23 | 0.527     | 0.997 | 0.895 | 0.994 |
| B | 19 | 0.575     | 0.921 | 0.691 | 0.248 |

arXiv: November 16, 2011

| | | | | | |
|---|---|---|---|---|---|
| C | 76 | 0.295 | 0.120 | 0.874 | 0.935 |

[1]Hoel (1986, pp 361)

Class A can be considered a separate group because the three correlation coefficients are greater than the critical coefficient. In class B the correlation coefficient of the function of surface gravity is below the critical coefficient, but in the functions of the radius and density, the correlation is higher than the critical coefficient. These two correlations justify considering class B as a separate group.

For its part, the class C, although the correlation coefficient in the function of the radius is very low for, in density and surface gravity are very high and justify separating the planets of this group as a class apart from the others.

To explain function (1) in terms of first principles we can start with the definition of density $\rho$

$$\rho = \frac{M}{V} \quad (6)$$

Where M is the planet mass and V is its volume. Supposing a spherical planet

$$\rho = \frac{M}{\frac{4}{3}\pi R^3} \quad (7)$$

Solving with respect to R we obtain

$$R = \left[\frac{3}{4\pi\rho}\right]^{\frac{1}{3}} M^{\frac{1}{3}} \quad (8)$$

On the other hand, the fitted equation for all the planets is:

arXiv: November 16, 2011

$$R = 3.42 \times 10^{-2} M^{0.3428} \qquad (1)$$

We can think that, in principle, equations (8) and (1) are the same and equation (1) is no more than an expression of (8) but in reality important differences between both of them exist.

On the one hand, to calculate R with equation (8) two variables, ρ and M, are needed, meaning that equation (8) is a function of two independent variables, and equation (1) is a function of only one independent variable.

Another difference between both equations is that if ρ and M are known, with equation (8), R can be precisely calculated for any planet. By contrast, with equation (1) an approximate value is obtained which must be corrected by a term o(M) which tell us the difference between the power law and the real value. But despite the fact that equation (8) gives a more precise value, the equation itself is useless because to know ρ, we need to know R and M, and if we know R then the equation (8) is not needed. On the other hand, with equation (1) we only need M to estimate R, with which we could calculate the radius of any planet from which we only know the mass, and this is the case of the exoplanets that are not of transit. Another important difference between equations (8) and (1) is that the coefficient of $M^{1/3}$ in the equation (8) is a function of ρ and the coefficient of $M^{0.3428}$ in the equation (1) is a constant.

Supposing that that both coefficients are the same, then

$$3.42 \times 10^{-2} = \left[\frac{3}{4\pi\rho}\right]^{1/3} \qquad (9)$$

And solving respect to ρ

$$\rho = \frac{3}{4\pi(3.42 \times 10^{-2})^3} \qquad (10)$$

And doing calculations

$$\rho = 5968 \, kg/m^3 \qquad (11)$$

This is a mysterious value since it is not the median density (1912 kg/m$^3$), the minimum density (115 kg/m$^3$), or the maximum density (26077 kg/m$^3$), none of the planets used to obtain equation (1) have this density. Therefore we can not say that it is the dominant value in density. Hence, it cannot be said that function (1) reflects function (8) and the fact that both have a similar exponent and a similar form is nothing more than a coincidence.

Analysis of the data clearly shows the existence of three classes of planets defined by the power laws for the radius, the density and the surface gravity as a function of mass. The transition from class A to class B is well definite. Both super-earths CoRoT-7b, the most massive planet of class A, and GJ 1214 b, the less massive planet of class B; have a mass of the same order of magnitude, 2.87x10$^{25}$kg for the first and 3.4x10$^{25}$kg for the second. But with respect to the density they are very different. CoRoT-7b has a density of 5549 kg/m$^3$ and GJ 1214 b has a density of 1576 kg/m$^3$. This transition is important since it clearly states the limit between the rocky planets, with a solid surface and geology, and the gas planets, without a solid surface and without geology.

The transition between class B and class C is less definite. The change in the tendency of the density is clear (figure 3). But planets are very concentrated around 10$^{27}$kg, then in the vicinity of this transition, it is not clear which planet belongs to which group. However, given the trends in both density and gravity it is clear that there is a boundary between the behavior of one group and another.

Two unusual facts were obtained: the density in class B decreases with increasing mass, and the radius of group C remains constant even though the mass increases. These two unexpected behavior can be explained as follows:

For some value of the mass in the vicinity of $3\times10^{25}$kg the surface escape velocity of the planets is high enough for the planet begins to accumulate a larger amount of light volatiles ($H_2O$, $CH_4$, H and He). As these gases are more abundant than the rest of the materials in the accretion disk, from there the main increase in mass, be for these gases. Because they are light gases, the volume will increase much quicker than the mass and so the density will decrease. This coincides with some authors models (Stevenson and Salpeter, 1976; de Pater, 2001, pg 244; Swift et al., 2011).

On the other hand, in Class C planets, the internal pressure is large enough so that the gases in its interior transform to much more dense phases, having a much lesser volume, so the radius will not change basically. This coincides with a maximum in the R (M) function which predicts certain models (Stevenson and Salpeter, 1976; de Pater, 2001, pg 244; Swift et al., 2011).

Another result which may seem unusual is that the surface gravity of 10 class B planets and 11 class C planets is less than the surface gravity of the Earth and they are planets with a mass between one and 3 orders of magnitude greater than that of the Earth. With gas giants it is extraordinary that they can accumulate enough hydrogen with such a weak gravity. This can be understood with a simple planetary model (toy model) with a nucleus with a mass of $3\times10^{25}$kg and a mean density of 6168 km/m$^3$ (maximum limit for group A) and a gaseous hydrogen covering with a constant density (300 km/m$^3$). Figure 5 shows the results of this model. The surface gravity decreases at first but then it increases, but it maintains itself below the terrestrial value. The escape velocity decreases a little above the transition mass between group A and group B but then it increases in a continuous way. During the mass interval of class B it remains above the terrestrial value. This explains the reason why the volatiles do not escape even though there is little gravity.

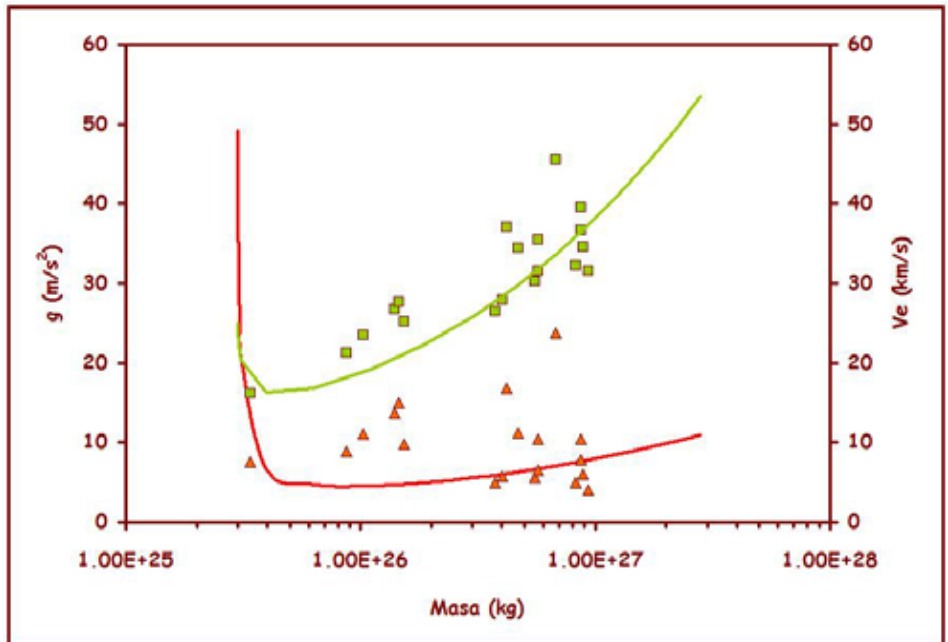

Figure 5 Toy model for the surface gravity (orange line) and the escape velocity (green line) of a planet with a nucleus of a mass of $3 \times 10^{25}$kg and a density of 6168kg/m$^3$ and a hydrogen gas covering at a constant density of 300 kg/m3 (this value was the one that best fit the data). Compared to the empirical surface gravity values (orange triangles) and the escape velocity (green squares).

The most relevant conclusions of this paper are:
The planets radius as well as the mean planet density and the surface gravity are mass power laws.
The planets, in the mass interval from $10^{19}$ to $10^{29}$kg, are clearly separated into three groups or classes of planets.
The A-class planets, which Earth and nearly all the planets of the Solar System belong, are in the mass interval of $10^{19} < M < 3 \times 10^{25}$kg, and the radius, density and surface gravity increase when the mass increases.
The B-class planets, which Saturn, Uranus and Neptune belong, are within the mass interval $3 \times 10^{25}$kg $< M < 10^{27}$kg. The radius increases with the mass and have the unexpected property of decreasing their density and their surface gravity when increasing their mass.

arXiv: November 16, 2011

The C-class planets, to which Jupiter belongs, are in the mass interval $10^{27} < M < 10^{29}$ kg and have the property that their radius stay constant while increasing their mass.

References

Basri, G. (2003). What is a planet?. Mercury. Nov./Dec. And on Internet: http://astro.berkeley.edu/~basri/defineplanet/Mercury.htm

De Pater, I. and J.J. Lissauer (2001) Planetary Sciences. Cambridge University Press.

Hoel, P.G. (1986) Estadística Elemental. Quinta Impresión. Editorial C.E.C.S.A.

Melosh, H.J. (2011). *Planetary Surface Processes*. Cambridge Planetary Science. Cambridge University Press.

Schneider, J. 2010. The Extrasolar Planets Encyclopedia, http://exoplanet.eu/

Stevenson, D.J. and E.E. Salpeter. (1976). Interior models of Jupiter. In *Jupiter*. Eds. T. Gehrels and M. S. Matthews. University of Arizona Press, Tucson pp 85-112.

Swift, D.C., J. Eggert, D.G. Hicks, S. Hamel, K. Caspersen, E. Schwegler, G.W. Collins, N. Nettelmann and G.J. Ackland. (2011). Mass-radius relationships for exoplanets. arXiv: 1001.4851v2

Tholen, D.J., V.G. Tejfel and A.N. Cox. 2000. Chapter 12 Planets and Satellites. In Allen's Astrophysical Quantities. 4$^{th}$ Edition Editor A.N. Cox. AIP Press and Springer.

Valencia, D. D.D. Sasselov, and R.J. O'Connell. (2007) Detailed Models of súper-Earths: How well can we infer bulk properties? Ap. J. 665: 1413-1420